\documentclass[10pt]{article}
\usepackage{amsmath}
\usepackage{graphicx}
\usepackage{amsfonts}
\usepackage{amssymb}
\usepackage{epsf}
\usepackage{latexsym}

\topmargin + 10pt

\def\d{\mbox{d}}

\def\md{\mbox{\scriptsize d}}

\def\sN{\mbox{\scriptsize N}}

\def\NSI{Na\"{\i}ve Schr\"{o}dinger Interpretation }
\def\CPI{Conditional Probabilities Interpretation }
\def\be{\begin{equation}}
\def\ee{\end{equation}}
\def\bea{\begin{eqnarray}}
\def\eea{\end{eqnarray}}
\def\fn{\footnote}

\def\d{\textrm{d}}

\def\cr{\mbox{\scriptsize{\bf $\mbox{ } \times \mbox{ }$}}}

\def\sB{\mbox{\scriptsize B}}

\def\fE{\mbox{\sffamily E}}
\def\fH{\mbox{\sffamily H}}

\def\fL{{\cal L}}

\def\fP{\mbox{\sffamily P}}
\def\fQ{\mbox{\sffamily Q}}
\def\fR{\mbox{\sffamily R}}
\def\fS{\mbox{\sffamily S}}
\def\fT{\mbox{\sffamily T}}
\def\fU{\mbox{\sffamily U}}
\def\fV{\mbox{\sffamily V}}

\begin{document}

%\begin{titlepage}

\begin{center}

{\Large{\bf SEMINAR ON RECORDS THEORY}} 

\vspace{.1in}

{{\bf Edward Anderson}}$^1$

\vspace{.1in}

{\large Peterhouse and DAMTP, Cambridge.}

\end{center}

%===========================================================ABSTRACT=======================================================================
\begin{abstract}

In quantum gravity, one seeks to combine quantum mechanics and general relativity.  
In attempting to do so, one comes across the `problem of time' impasse: 
the notion of time is conceptually different in each of these theories.  
In this seminar, I consider the timeless records approach toward resolving this.  
Records are localized, information-containing subconfigurations of a single instant.
Records theory is the study of these and of how science (or history) is to be abstracted from 
correlations between them.  
I explain how to motivate this approach, provide a ground-level structure for it and 
discuss what kind of further tools are needed.  
For a more comprehensive account with many more references, see \cite{arxiv}.  

\end{abstract}

%==========================================================================================================================================

%\noindent$^1$ ea212@cam.ac.uk
%
%\end{titlepage}

%==========================================================================================================================================
%==========================================================================================================================================
%==========================================================================================================================================
\section{Introduction}\label{Intro}
%==========================================================================================================================================
%==========================================================================================================================================
%==========================================================================================================================================

Records play a role in Quantum Cosmology and Quantum Gravity.  
The literature on this is a heterogeneous, consisting of    
1) reinterpretations \cite{Bell, B94II} of how $\alpha$-particle tracks form in a bubble chamber 
\cite{Mott} that may be analogous to Quantum Cosmology \cite{HalliMott, Hallioverlap, 
CaLa, CastAsym}.   
Therein, Barbour's approach also involves reformulating classical physics in timeless terms \cite{BB82, 
B94I, RWR, Phil} and places emphasis on the configuration of the universe as a whole and on 
timelessness casting mystery \cite{B94II} upon why `ordinary physics' works.     
2) The {\it\CPI} for Quantum Cosmology \cite{CPI, PAOT, Page} places its emphasis on subconfigurations 
(SC's) of the universe within a single instant.  
Here, ordinary physics of SC's ends up familiarly explained through other SC's providing approximate 
time standards for them, and what is habitually observed is the dynamics of subsystems rather than of 
the whole universe \cite{BS, B94I, B94II, 06I}.    
3) While Histories Theory (see e.g. \cite{GMH}) is not primarily timeless, a Records Theory sits within 
it \cite{GMH, H99}, and benefits from inheriting part of the structural framework developed for 
Histories Theory.

\mbox{ }

\noindent \hspace{0.4in} Records are ``{\it somewhere in the universe that information} 

\noindent \hspace{0.4in} {\it is stored when histories decohere}" (p 3353 of \cite{GMH}) $\mbox{ } . 
\hspace{0.4in} \mbox{ (0)}$   

\mbox{ } 

This seminar's Records Theory is a synthesis of elements drawn from 1 

\noindent to 3.  
In outline, I consider {\it records} to be information-containing SC's of a single instant.  
{\it Records Theory} is then the study of these and how dynamics (or history or science) is to be 
abstracted from correlations between same-instant records.  
It is to make this abstraction meaningful that I insist on records being SC's rather than whole instants, 
thus getting round the abovementioned `mystery' by a similar argument to the Conditional 
Probability Interpretation's.

For adopting a Records Theory approach to profitable, I argue that records should have the following 
properties.   
{\it Useability}, in that A) their whereabouts [c.f. (0)] should be spatially-localized SC's of the 
universe, for whatever notion of space that one's theory has and restricted to the observationally 
accessible part thereof.  
B) They should also belong to a part of the SC space for which observational imprecisions in 
identifying SC's do not distort the extraction of information too much  
{\it Usefulness}: their information content [c.f. (0)] should be high enough and of the 
right sort of quality to enable reliable measures of correlation to be computed.
Not all systems have instants solely of this nature, so Records Theory may not always be profitable.  
1) and 2) additionally require {\it semblance of dynamics} to emerge from timeless records.

Sec 2 summarizes motivation for Records Theory.  
Sec 3 mentions some illustrative toy models. 
Sec 4 proposes a ground-level structure for Records Theory which parallels some of that of Histories 
Theory.  
I then comment on the useability, usefulness and correlation aspects of records in Sec 4--6, 
and more speculative aspects in Sec 7.

%========================================================================================================
%==========================================================================================================================================
%==========================================================================================================================================
\section{Some motivations for Records Theory}\label{Motivation}
%==========================================================================================================================================
%==========================================================================================================================================
%========================================================================================================

Records Theory should be motivated as follows \cite{arxiv}.  
1) The Problem of Time (POT) in Quantum Gravity is an incompatibility between the roles played by `time' 
in GR and in QM \cite{Isham93}.  
One conceptually clear way of dealing with this problem is to recast both GR and QM in a timeless mold.  
[While the Wheeler--DeWitt equation (WDE)'s timelessness might specifically prompt some physicists toward 
Timeless Records Theory, this equation has numerous technical problems and may not be trustworthy.   
Despite e.g. \cite{BS, B94I, B94II}, nor should one turn to Timeless Records Theory due to earlier detailed documentation of problems 
with the other POT approaches, but rather judge it due to its own merits and shortcomings (Sec 7).] 
2) One can in principle treat all of change, processes, dynamics, history and the scientific enterprise 
in these timeless terms.  
The classification and subsequent partial elimination of question types in this seminar is to demonstrate 
that (and its impracticality in some cases).  
3) Records Theory is of potential use in removing some unclarities (see e.g. \cite{Hallioverlap}) from 
the foundations of Quantum Cosmology (which in turn is what Inflationary Theory is to rest on).  
%
%Via the above contact with testable assertions, hitherto philosophical contentions about QM 
%(in particular as applied to closed systems such as the whole universe) may enter into mainstream physics.     
%
Records Theory makes contact with this area in e.g. the following ways.  
\noindent A) Such as CMB inhomogeneities or the pattern and spectra of galaxies may be considered to be
useful records.  
\noindent B) Within a histories perspective, the decoherence process makes records, but information is 
in general lost in the making.  
E.g. mixed states necessarily produce imperfect records \cite{H99}.   
Furthermore, finding out where in the universe the information resides (i.e. where the records are) 
should be capable of resolving in which cases gravity decoheres matter or vice versa.  
Decoherence is habitually linked with the emergence of (semi)classicality, so there may well be 
some bridge between Records Theory and the Semiclassical Approach.  
\noindent C) See also Sec 7.  
4) Records Theory is (alongside Histories Theory) a universal scheme in that all types of theory or  
system admit a such.

It has also been suggested that records are more operationally meaningful than the histories.  
For, study of records is how one does science (and history) in practise? 
Unfortunately, this last suggestion fails {\sl as motivation}, because of the difference between the notion of records 
as in (some of the) SC's that the system provides and as in things which are localized, accessible and 
of significant information content.  
As effective reconstruction of history requires the SC's in question to have these in general 
unestablished properties, it is a {\sl question to address} rather than a preliminary motivation to have 
that records are more operationally meaningful than histories through possessing good enough qualities 
to permit a meaningful such reconstruction.    
Thus what one should do is 1) pin down where the ``somewhere" in (0) is (the central motivation in some 
of Halliwell's papers, e.g. \cite{H99}).    
2) Determine whether the record thereat is useful --  Gell-Mann--Hartle assert that what they call 
records ``{\it may not represent records in the usual sense of being constructed from quasiclassical 
variables accessible to us}" (p 3353 of \cite{GMH}).  
Also, it may be that the $\alpha$-particle track in the bubble chamber is atypical in its neatness and 
localization.  
For, bubble chambers are carefully selected environments for revealing tracks -- much human trial and 
error has gone into finding a piece of apparatus that does just that.  
$\alpha$-tracks being useful records could then hinge on this careful pre-selection, records in general 
then being expected to be (far) poorer, as suggested e.g. by Joos--Zeh's paradigm \cite{JZ} of a dust 
particle decohering due to the microwave background photons.  
In this situation, records are exceedingly diffuse as the information is spread around by the CMB 
photons.

%========================================================================================================
%==========================================================================================================================================
%==========================================================================================================================================
\section{Toy models for Records Theory}\label{Toys}
%==========================================================================================================================================
%==========================================================================================================================================
%========================================================================================================

Ordinary (conservative) mechanics already has a simple analogue of the 
Hamiltonian constraint: 
a homogeneous quadratic energy constraint which 
gives a time-independent Schr\"{o}dinger equation 
(TISE) analogue of the WDE at the quantum level.
There are furthemore other mechanics that share more features with GR: relational particle mechanics 
(RPM's) are such \cite{BB82, 06I, Triangle, 07II}.  
These have additional linear constraints [such as a zero angular momentum constraint, 
the physics then being encoded solely in the relative separations and angles rather than in any 
absolute angles, in analogy with how the linear momentum constraint of GR is interpretable 
in terms of the physics being in the shape of 3-space and not in its coordinatization.]  
Scale-invariant RPM's additionally have a linear zero dilational constraint 
that is analogous to the maximal slicing condition in GR.  
Full reductions are available \cite{Triangle} for 2d RPM's (of which the scale-invariant one is better 
behaved) allowing us to do quite a lot more with these particular models.  
The kinetic term then contains the positive-definite Fubini--Study metric.

One could include a harmonic oscillator detector within one's mechanics model, or 
couple it to an up--down detector.  
These can hold information about one Fourier mode in the signal, thus showing that even very simple 
systems can make imperfect records \cite{H99}.  
Finally, the inhomogeneous perturbations about homogeneous cosmologies \cite{HallHaw} are 
a more advanced toy model (with which RPM's nevertheless share various features).

%==============================================================================================================
%==============================================================================================================
%==========================================================================================================================================
\section{SC space structure and useable records}\label{GroundLevel}
%==========================================================================================================================================
%==============================================================================================================
%==============================================================================================================

%==============================================================================================================
%==============================================================================================================
\subsection{First level of classical structure}
%==============================================================================================================
%==============================================================================================================

A {\it configuration} $Q_{\Delta}(p)$ is a set of particle positions and/or field values, where $\Delta$ 
is a multi-index which covers both particle and field species labels and whatever `tensorial' indices each 
of these may carry, and $p$ is a fixed label.

Hierarchical, nonunique splittings into subsystems can then be construed: 
$Q_{\Gamma}(p)$ is a subsystem of $Q_{\Delta}(p)$ if $\Gamma$ is a subset of the indexing set $\Delta$.  
The {\it finest} such subdivision is into individual degrees of freedom.

Two question-types that may be considered at this level are: 
\noindent {\bf Be$1^{\prime}$}), does $Q_{\Delta}(p)$ have acceptable properties? (That covers 
both mathematical consistency and physical reasonableness).  
\noindent {\bf Be$2^{\prime}$}) If properties of $Q_{\Delta}(p)$ are known, does this permit deduction of 
any observable properties of some $Q_{\Delta^{\prime}}(p)$ for $\Delta^{\prime}$ disjoint from $\Delta$? 
In other words, are there observable correlations between SC's of a single instant?

%%%%%%%%%%%%%%%%%%%%%%%%%%%%%%%%%%%%%%%%%%%%%%%%%%%%%%%%%%%%%%%%%%%%%%%%%%%%%%%%%%%%%%%%%%%%%%%%%%%%%%%%%
%%%%%%%%%%%%%%%%%%%%%%%%%%%%%%%%%%%%%%%%%%CONFIGURATION SPACES%%%%%%%%%%%%%%%%%%%%%%%%%%%%%%%%%%%%%%%%%%%

Many notions and constructions that theoretical physicists use (see e.g. 

\noindent \cite{Isham93}) 
%
%p 15
%
additionally require consideration of sets of instants.  
A {\it configuration space of instants} is  $\fQ_{\Delta} = \{ Q_{\Delta}(p): \mbox{ } p$ 
a label running over a (generally stratified) manifold $\}$.  
This is a {\it heap} of instants.  
One defines SC spaces similarly.
The counterpart of decomposition into subsystems is now a break-down into subspaces.  
An ordinary (absolute) particle mechanics configuration space 
is the set of possible 
positions of N particles, $\fQ$(N, d).  
Relational configuration space ${\cal R}$(N, d) is the set of possible relative separations and relative 
angles between particles.  
Preshape space $\fP$(N, d) is the set of possible scale-free relative particle positions.
Shape space $\fS$(N, d) is the set of possible scale-free relational configurations.
As another example, for geometrodynamics, a rather redundant configuration space is Riem($\Sigma$): the 
space of positive-definite 3-metrics $h_{\mu\nu}(x_{\gamma})$ on the 3-space of fixed topology $\Sigma$.  
Less redundant ones are superspace($\Sigma$) = Riem($\Sigma$)/Diff($\Sigma$) and  
(something like) conformal superspace($\Sigma$) = Riem($\Sigma$)/Diff($\Sigma$) $ \times $ 
Conf($\Sigma$), 
%(see e.g. \cite{ABFKO}), 
for Diff($\Sigma$) the diffeomorphisms of $\Sigma$ and Conf($\Sigma$) the conformal transformations of 
$\Sigma$.

While each $\fQ_{\Delta}$ corresponds to a given model with a fixed list of contents, one may not know 
which model a given (e.g. observed) SC belongs to, or the theory may admit operations that alter the 
list of contents of the universe.    
Then one has a {\it grand heap} of SC spaces of instants.  
For example, use 1) $\bigcup_{\sN \in \mbox{ \scriptsize $\mathbb{N}_0$ }}\fQ$(N, d) for 
a mechanics theory that allows for variable particle number.
2) $\bigcup_{\mbox{ \scriptsize various $\Sigma$}}$ superspace($\Sigma$) 
for a formulation of GR that allows for spatial topology change.

A second type of hierarchical splitting are {\it grainings}: the various ways that $\fQ$ can be 
partitioned. 
These define a partial order $\prec$ on the subsets of $\fQ$.  
$A \prec B$ is termed `$A$ is finer grained than $B$', while 
$C \succ D$ is termed `$C$ is coarser-grained than $D$'.  
The coarsest grained set is $\fQ$ itself, while the finest grained sets are each  
individual $q(p)$ (the constituent points of $\fQ$).

{\sl Localization in space} continues to be formulable in the relational context \cite{arxiv}.  
{\sl Localization on configuration spaces} \cite{arxiv, ARec2} can sometimes be attained by augmenting the configuration 
space to be equipped with a norm.  
E.g. on $\fQ$(N, d), these are the obvious unweighted and (inverse) mass-weighted 
$\mathbb{R}^{\sN\md}$ norms, which still play a role in more reduced 
configuration spaces through these `inheriting' structures such as the $\mathbb{R}^{nd}$ norm for  
$\fR$(N, d) or the chordal norm for $\fP$(N, d).
%\fn{For 
%%%%%%%%%%%%%%%%%%%%%%%%%%%%%%%%%%%%%%%%%%%%%%%%%%%%%%%%%%%%%%%%%%%%%%%%%%%%%%%%%%%%%%%%%%%%%%%%%%%%%%%%%
%all that such spaces are termed relational, they still bear imprints of the absolute.  
%
%Other examples are a residual sense of dimensionality and some topological aspects being inherited  
%\cite{FORD}.}  
%%%%%%%%%%%%%%%%%%%%%%%%%%%%%%%%%%%%%%%%%%%%%%%%%%%%%%%%%%%%%%%%%%%%%%%%%%%%%%%%%%%%%%%%%%%%%%%%%%%%%%%%%
%
If the configuration space has a natural metric more complicated than the Euclidean one, one 
might be able to extend the above notion to the norm corresponding to that.  
E.g., one could use the Fubini--Study norm on $\fS$(N, 2), or the inverse DeWitt line element on 
Riem($\Sigma$) (but its indefiniteness causes some problems).

Another way is to intrinsically compute on each configuration a finite number of quantities, 
i: $\fQ \longrightarrow \mathbb{R}^n$, and then use the $\mathbb{R}^n$ norm 
$D_{\mbox{\scriptsize Eucl}}^{\mbox{\scriptsize i}}(Q_{\Delta}, Q_{\Delta}^{\prime})$ $= 
||\mbox{i}(Q_{\Delta}) - \mbox{i}(Q_{\Delta}^{\prime})||^2$
(though this is limited for some purposes by i having a nontrivial kernel).
E.g. one can compare SC's in $\fQ$(N, d), $\fR$(N, d) or ${\cal R}$(N, d) by 
letting i be the total moment of inertia for each SC (a mass-weighted norm).  
In geometrodynamical theories, one could additionally compute geometrical quantities to serve as i, or 
embed $N$ points in a uniformally random way in each geometry and then use the pairwise metric 
distances between the points to furbish a vectorial i.  
Or, one could use total volume, anisotropy parameter or a vector made out of these, or use curvature 
invariants such as maximal or average curvatures of a given 3-space (e.g. objects related to the Weyl 
tensor which are also perported measures of gravitational information, see Sec 5).      
Or, for nonhomogeneous GR, one could compute eigenvalues of an operator D associated with that geometry 
(e.g. Laplacian or Yano--Bochner operators)
and construct a spectral measure i from these.
Another measure of inhomogeneity that could be used as an i would be an energy density contrast 
type quantity F[$\varepsilon/\langle\varepsilon\rangle]$ (for $\varepsilon$ the energy density 
distribution and $\langle \mbox{ } \rangle$ denoting average over some volume) such as 
$\varepsilon/\langle\varepsilon\rangle$ or 
$\left\langle
\frac{\varepsilon}{\langle\varepsilon\rangle}\mbox{log}
\left(
\frac{\varepsilon}{\langle\varepsilon\rangle}
\right) 
\right\rangle$,
which particular functional form \cite{HBM} also has information content connotations (see Sec 6). 
One can readily supply a notion of `within $\epsilon$ of' for each above structure (contingent 
to what distance axioms it obeys), thus obtaining examples of grainings.  
RPM's with their local particle clusters, and inhomogeneous perturbations about minisuperspace 
with their localized bumps, are two such settings.

Four further question types can then be addressed.  
Two generalize their primed counterparts to model the imperfection of observation.  
\noindent {\bf Be$1$}), does $q_{\Delta}(P)$ have acceptable properties? 
This is now for a graining set P rather than for an individual instant p.    
\noindent {\bf Be$2$}), if properties of $q_{\Delta}(P)$ are known, does this permit deduction of any 
properties of $q_{\Delta^{\prime}}(P)$ for $\Delta^{\prime}$ disjoint from $\Delta$?
\noindent The other two involve the $\fQ$ space of the theory or theories that the observations are 
perported to belong to.  
\noindent
{\bf BeS$1$}) is: what is $\mbox{P}(q_{\Delta}(P))$ within the collection of SC spaces? 
\noindent
{\bf BeS$2$}) is: what is P($q_{\Delta^{\prime}}(P)$ has properties ${\cal P}^{\prime} |  q_{\Delta}(P)$ has 
properties ${\cal P}$)?\footnote{P denotes probability and 
%%%%%%%%%%%%%%%%%%%%%%%%%%%%%%%%%%%%%%%%%%%%%%%%%%%%%%%%%%%%%%%%%%%%%%%%%%%%%%%%%%%%%%%%%%%%%%%%%%%%%%%%%
$|$ denotes `given that', i.e. conditional probability.}
%%%%%%%%%%%%%%%%%%%%%%%%%%%%%%%%%%%%%%%%%%%%%%%%%%%%%%%%%%%%%%%%%%%%%%%%%%%%%%%%%%%%%%%%%%%%%%%%%%%%%%%%%
%
\noindent Examples of such questions are: what is P(space is almost flat)? 
What is P(space is almost isotropic)? 
What is P(space is almost homogeneous)?

%=======================================================================================================
%=======================================================================================================
\subsection{Configuration comparers and decorated instants}
%=======================================================================================================
%=======================================================================================================

The above single-configuration notion of closeness may not suffice for some 

\noindent purposes (whether in principle or through lack of mathematical structure 
leaving one bereft of theorems through which to make progress).  
Other notions of closeness on the collection may depend on a fuller notion of comparison {\sl between} 
instants, i.e. their joint consideration rather than a subsequent comparison of real numbers extracted 
from each individually.   
That may serve as a means of judging which instants are similar, or of 
which instants can evolve into each other along dynamical trajectories.  
Some criteria to determine which notion to use \cite{ARec2} are adherence to the axioms of distance, 
gauge or 3-Diffeomorphism invariance as suitable, and, for some applications, whether it can be applied 
to grand heaps.

One way of providing comparers is to upgrade the previous subsection's normed spaces and geometries to inner 
product spaces, metric spaces and topological spaces \cite{Triangle, arxiv}.  
In the case of inner products or metrics, ${\cal M}^{\Gamma\Delta}Q_{\Gamma}Q^{\prime}_{\Delta}$ then 
supplies a primitive comparer of unprimed and primed objects $Q_{\Gamma}$, $Q_{\Delta}^{\prime}$.

Also consider replacing $\fQ_{\Delta}$ by 
the tangent bundle $\fT(\fQ_{\Delta})$ (configuration-velocity space \cite{B94I}), or 
the unit 

\noindent
tangent bundle $\fT_u(\fQ_{\Delta})$ (configuration-direction space), or 
the cotangent bundle $\fT^*(\fQ_{\Delta})$ (configuration-momentum space, which, 
if augmented by a symplectic structure, is phase space). 
Such notions continue to exist for restricted configuration spaces in cases with constraints.  
This last feature involves quotienting operations, which can considerably complicate structure in 
practise.    
Envisage all these as {\it `heaps of decorated instants'}, $\fH$, which more general notion I use 
to supercede $\fQ$.

A common situation is to compare not configurations $Q_{\Gamma}$ and $Q_{\Delta}^{\prime}$ but rather 
the corresponding velocities $\dot{Q}_{\Gamma}$ and $\dot{Q}_{\Delta}^{\prime}$, with the comparer 
employing the kinetic metric 
An example of such a comparer is the Lagrangian  
$\fL: \fT(\mbox{G-bundle over } \fQ) \longrightarrow \mathbb{R}$ 
$
\fL[Q_{\Delta}, g_{\Lambda}, \dot{Q}_{\Delta}, \dot{g}_{\Lambda}] = 2\sqrt{\fT\{\fU + \fE\}} \mbox{ } ,  
\label{Lag}
$
where, in this seminar's examples, $\fU$ is minus the potential term $\fV(Q_{\Delta})$ and 
$\fT$ is the kinetic term 
$\fT[Q_{\Delta}, g_{\Lambda}, \dot{Q}_{\Delta}, \dot{g}_{\Lambda}] = 
M^{\Gamma\Delta}(Q_{\Theta})\{\stackrel{\longrightarrow}{G}\dot{Q}_{\Gamma}\}
                                         \stackrel{\longrightarrow}{G}\dot{Q}_{\Delta}/2$
for $\stackrel{\longrightarrow}{G}$ the action of the group G of redundant motions whose generators are 
parametrized by auxiliary variables $g_{\Lambda}$.  
[Here, the dot denotes the derivative with respect to label-time, an overall time that is meaningless 
because the actions considered are invariant under label change (= reparametrization)].

%%%%%%%%%%%%%%%%%%%%%%%%%%%%%%%%%%ACTION AS COMPARER AND OTHER ACTIONS%%%%%%%%%%%%%%%%%%%%%%%%%%%%%%%%%%%
%
This also exemplifies that one often corrects the $Q_{\Gamma}$ or $\dot{Q}_{\Gamma}$ with respect to a 
group G of transformations under which they are held to be physically unchanged.  
That involves the group action of G on the $Q_{\Gamma}$ or $\dot{Q}_{\Gamma}$.  
E.g. for particle velocities $\dot{q}_{i\alpha}$, 
%the infinitesimal action of the translations (generated by $a_{\alpha}$) is 
%$\dot{q}_{i\alpha} \longrightarrow \mbox{ } \stackrel{\rightarrow}{T} \dot{q}_{i\alpha} = \dot{q}_{i\alpha} + 
%\dot{a}_{\alpha}$, 
the infinitesimal action of the rotations (generated by $b_{\alpha}$) is 
$\dot{q}_{i\alpha} \longrightarrow \mbox{ } \stackrel{\rightarrow}{R} \dot{q}_{i\alpha} = \dot{q}_{i\alpha} + 
q_{i\alpha} \cr \dot{b}_{\alpha}$   
%and the infinitesimal action of the dilations (generated by c) is 
%$\dot{q}_{i\alpha} \longrightarrow \mbox{ } \stackrel{\rightarrow}{D} \dot{q}_{i\alpha} = \dot{q}_{i\alpha} + 
%\dot{c} q_{i\alpha}$ .  
%
E.g. for 3-metric velocities $\dot{h}_{\mu\nu}$, 
the infinitesimal action of the 3-diffeomorphisms (generated by $B_{\alpha}$) is 
$\dot{h}_{\mu\nu} \longrightarrow \mbox{ } \stackrel{\rightarrow}{\mbox{Diff}} \dot{h}_{\mu\nu} = 
\dot{h}_{\mu\nu} - \pounds_{\dot{B}}h_{\mu\nu}$.   
One furthermore often then minimizes with respect to the group generator (arbitrary frame  
`shuffling auxiliary').  
This ensures the physical requirement of G-invariance (i.e. gauge invariance, including 3-diffeomorphism 
invariance in geometrodynamics).

Then one has e.g.   
1) The {\it Kendall-type comparer} \cite{arxiv}  
$\stackrel{\mbox{\scriptsize min}}{\mbox{\scriptsize $g \in$ G}} 
{\cal M}^{\Gamma\Delta}{\cal Q}_{\Gamma}\stackrel{\rightarrow}{G} {\cal Q}^{\prime}_{\Delta}$  
for $\stackrel{\rightarrow}{G}$ the {\sl finite} group action.  
2) Construct 
$
{\cal M}^{\Gamma\Delta}\{\stackrel{\rightarrow}{G} \dot{Q}_{\Gamma}\}
                         \stackrel{\rightarrow}{G} \dot{Q}_{\Delta}
$
for $\stackrel{\rightarrow}{G}$ the {\sl infinitesimal} group action.  
Then weight by $\fU + \fE$, square-root, then integrate with 
respect to spatial extent if required and with respect to label time so as to produce the corresponding 
action.  
Variation of this ensures G-independence.  
Actions of this form include \cite{arxiv} the Jacobi action for mechanics, Barbour--Bertotti type 
actions for RPM, and the Baierlein--Sharp--Wheeler type actions for geometrodynamics.    
The variational procedure then entails minimization with respect to $g_{\Lambda}$.
One could also weight by $1/\{\fU + \fE\}$ and square-root.  
This gives Leibniz--Mach--Barbour timefunctions (c.f. \cite{B94I, SemiclI}).  
Another variant is the DeWitt measure of distance: let one $\dot{h}_{\alpha\beta}$ 
and 1 of the 2 metrics in each factor of the DeWitt supermetric be with respect to primed coordinates, 
integrate with respect to both primed and unprimed space, and {\sl then} square-root.   
One then obtains a semi-Riemannian metric functional (in the sense of `Finslerian metric function').  
In inhomogenous geometrodynamics, one can likewise decompose combined measures of 
local size and shape into separate comparers (c.f. \cite{ARec2} and techniques in \cite{ABFKO}).   
In each case, the individual rather than combined comparers are better-behaved as distances.

Comparers along the lines of 1) and 2) are universal, insofar as they apply both to RPM's and to GR.  
However, the GR version has an indefinite inner product which does not confer 
good distance properties in contrast to the positive definite one in mechanics.  
Thus one might need different tools in each case, or use only the {\sl shape part} of the GR inner 
product, which is itself positive definite \cite{ARec2}.  
% 
%3) In the case of 3-metrics, another comparer whose Diff-independence is assured by a similar method to 
%the above is Gromov's distance between Riemannian spaces \cite{Gromov}.
%
%\mbox{ }
%
Instead of using highly redundant variables alongside gauge auxiliaries and a shuffling procedure, one 
could work with reduced gauge-invariant configurations $Q_{\Omega}$, for $\Omega$ a smaller indexing set 
than $\Delta$, and a Lagrangian $\widetilde{\fL}: \widetilde{\fT}(\fQ_{\Omega}) \longrightarrow 
\mathbb{R}$ constructed from these, 
$
\widetilde{\fL}[Q_{\Omega}, \dot{Q}_{\Omega}] = 
2\sqrt{        \widetilde{\fT}    \{    \widetilde{\fU} + \widetilde{\fE}    \}         }
$ 
for $\widetilde{\fT}[Q_{\Omega}, \dot{Q}_{\Omega}]$ a suitable, `more twisted' kinetic term.  
While, one seldom has this luxury of explicit gauge-invariant variables being available, it is   
available \cite{Triangle} for the 2d RPM of pure shape.   
The reduced configuration space metric is the Fubini--Study metric, from which this example's     
`more twisted' kinetic term is formed.  
The associated notion of distance is then useable between 2d shapes.
Alternatively, one could work with (more widely available) secondary quantities that are guaranteed to 
have the suitable invariances, e.g. further spectral measures.

If there's a sense of more than one instant, there is one becoming question type per question 
type above, by the construction Prob(if $Q_{\Delta}(p)$ has properties ${\cal P}$ then it becomes 
$Q^{\prime}_{\Delta}(p^{\prime})$ with properties ${\cal P}^{\prime}$).  
I denote each such question type as above but with `{\bf Become}' rather than `Be'.

%=======================================================================================================
%=======================================================================================================
\subsection{If there were a notion of time}
%=======================================================================================================
%=======================================================================================================

Then yet further question types would emerge.  
Each of the non-statespace 

\noindent
questions can now involve each instant being {\sl prescribed to be at 
a time}.  
I denote this by appending a {\bf T}.  
The new Be questions concern `being at a particular time',     
while the new Become questions are of the form `X at time 1 becomes Y at time 2'.  
For questions concerning heaps there is a further ambiguity:  
`at any time' now makes sense as well as `at a particular time'.   
Thus for each BeS question there are two {\bf BeST} questions (denoted {\bf a}, {\bf b}), 
and for each BecomeS question there are four {\bf BecomeST} questions (denoted {\bf a}, {\bf b}, {\bf c}, 
{\bf d}).  
Thus 32 question-types have been uncovered.  

%\mbox{ }
%
%\noindent 
%[{\bf Fig 1}.  The various question types and which moves remove some of them as separate entities.  
%  
%Single-SC questions (no S's) are at most consistency checks, while S-questions 
%inter-relate observations and are thereby {\sl testable} if there is another LCR to bring into the 
%picture.]   

%=======================================================================================================
%=======================================================================================================
\subsection{Further analysis of question-types and of time}
%=======================================================================================================
%=======================================================================================================

First note {\bf Suppression 1}: the 8 primed questions are clearly just subcases of their more realistic 
unprimed counterparts.  
Next note that the previous subsection crucially does not say what time is.  
Ordinary classical physics is easily excused: there is a real number valued external time, 
so that each $\fH$ is augmented to an extended heap space $\fH \times \mathbb{R}$.
One key lesson from GR, however, is that there is no such external time.    
Stationary spacetimes (including SR's Minkowski spacetime) do possess a timelike Killing vector, 
permitting a close analogue of external time to be used, but the generic GR solution permits no such 
construction.
The generic solution of GR has a vast family of coordinate timefunctions, none of which has a privileged 
status unlike that associated with a stationary spacetime's timelike Killing vector.    
Questions along the lines of those above which involve time need thus specify {\sl which} time.  
Using `just any' time comes with the multiple choice and functional evolution \cite{Isham93} 
subaspects of the POT -- this ambiguity tends to lead to inequivalent physics at the quantum level.

Another way of partly adhering to the above key lesson, which can be modelled at the level of 
nonrelativistic but temporally-relational mechanical models, is that `being, 
at a time $t_0$' is {\sl by itself} meaningless if one's theory is time label reparametrization 
invariant.

Alternatives that render particular times, whether uniquely or in families up to frame embedding variables, 
meaningful are specific internal, emergent or apparent time approaches.
Therein, time is but a property that can be read off the (decorated sub)configuration.  
E.g. the notion of time in \cite{CPI} can be thought of in this way.  
Thereby one has {\bf Subsumption 2}: all question types involving a T are turned into the corresponding 
question types without one.
%\fn{It is not clear which as in this setting one can have in principle 
%different configurations take the same time value (e.g. through lying on different paths of motion).}  
%
This property might concern a clock within the environment/background, within the subsystem under study, 
or partly within both. 
Indeed, one could have a universe-time to which all parts of the configuration contribute 
rather than a clock {\sl sub}system.

{\bf Subsumption 3}: Each BecomeST b, c pair becomes a single question type if there is time 
reversal invariance.  
{\bf Subsumption 4}: If the time used is globally defined on $\fH$, BeSTb questions and 
BecomeSTd questions are redundant.     
This can in any case be attained by considering restricted $\fH$ defined so that this is so.   
(Whether that excludes interesting physics is then pertinent).  
At this stage, one is left with 8 question types.

{\bf Subsumption 5} has been suggested by Page (e.g \cite{PAOT}) and also to some extent 
Barbour \cite{B94II}.   
%p 19
It consists in supplanting all becoming questions by more operationally accurate being questions as 
follows.    
It is not the past instant that is involved, but rather this appearing as a memory/subrecord in 
the present instant, alongside the subsystem itself.   
Thus this is in fact a correlation within the one instant.  
In this scheme, one does not have a sequence 
of events but rather one present event that contains memories or other evidence of `other events'.\fn{As 
an illustrative sketch, one can imagine a configuration in which the record actually under study is the 
na\"{i}ve record plus the observer next to it, whose memory includes a SC which encodes 
himself peering at the record `at an earlier time' and a SC in which he has this first 
memory and a prediction `derived from it'.}

If subsumption 5 is adopted, the remaining question types are Be2 about how likely a correlation between 
two subsystems within the one grained subinstant is, theory-observation question type BeS2 about how 
likely an instant is within a statespace, and two `consistency' question types Be1 and BeS1 about 
properties of a subinstant.  
If subsumption 5 is not adopted (or not adoptable in practise), there are additionally four corresponding 
types of becoming questions. 
Reasons why subsumption 5 might not be adopted, or might not be a complete catch-all of what one would 
like to be explained include 
I) impracticality: studying a subsystem S now involves studying a larger subsystem containing multiple 
imprints of S.
Models involving memories would be particularly difficult to handle (see footnote 2). 
II) If one wants a scheme that can explain the Arrow of Time, then Page's scheme looks to be 
unsatisfactory.  
While single instants such as that in footnote 2 could be used to simulate the scientific process as 
regards `becoming questions', N.B. that these single instants correspond to the {\sl latest} 
stage of the investigation (in the `becoming' interpretation), while `earlier instants' will not have 
this complete information.  
III) Additionally, important aspects of the scientific enterprise look to be incomplete in this approach 
-- in interpreting present correlations, one is in difficulty if one cannot affirm that one did in fact prime the measuring 
apparatus would appear to retain its importance.    
I.e. as well as the `last instant' playing an important role in the interpretation, initial conditions 
implicit in the `first instant' also look to play a role \cite{arxiv}.

%=======================================================================================================
%=======================================================================================================
\subsection{QM and beyond?}
%=======================================================================================================
%=======================================================================================================

At the classical level, one could either take certainty to be a subcase of probability, or note that 
even classically it is probabilities that are relevant in practise -- e.g. due to limits on precision of 
observations.    
2) A notion of $\mbox{P(trajectory goes through a subregion $\Delta$}$ 
for each space $\fH$) is then required.
This is particularly common in the literature in the case in which $\fH$ is phase space.   
Then if one canonically-quantizes, the Hamiltonian provides a TISE such as the WDE in the case of GR.

Because they refer to configurations, such as the almost flat, almost isotropic and almost homogeneous 
questions have obvious counterparts in configuration--representation QM; in concerning pieces of the 
configuration space these questions lie outside the usual domain of QM.  
The \NSI and the \CPI are two interpretations outside or beyond conventional QM formalism suggested 
to answer such questions.     
The former serves to address the BeS1 version of this paragraph's questions, such as 
what is P(universe is almost flat) or what is P(Inflation) \cite{arxiv}.  
The latter addresses Be2 or BeS2 questions such as 
P(One part of the sky is smooth $|$ another is) 
[all within a given instantaneous configuration].

%=====================================================================================================
%=====================================================================================================
%=====================================================================================================
\section{Are records typically useful?}
%=====================================================================================================
%=====================================================================================================
%=====================================================================================================

Records Theory requires A) for SC's to be capable of holding enough information to address 
whatever issues are under investigation.  
Thus Information Theory is pertinent.   
Information being (more or less) negentropy, a starting classical notion is the Boltzmann-like 
$
I_{\mbox{\scriptsize Boltzmann}} = - \mbox{log}W
\label{Boltz}
$
(using $k_{\sB} = 1$ units) for $W$ the number of microstates.  
One could furthermore use such as {\it Shannon information}, 
$
I_{\mbox{\scriptsize Shannon}}(p_x) = \sum_x p_x\mbox{log}p_x
$ 
for $p_x$ a discrete probability distribution for the records, or  
$
I_{\mbox{\scriptsize Shannon}}[\sigma] = \int \d\Omega \sigma\mbox{log}\sigma
\label{ShaCont}
$    
for $\sigma$ a continuous probability distribution.  
If one considers records at the quantum level, then one could instead use such as {\it von Neumann 
information}, 

\noindent
$I_{\mbox{\scriptsize von Neumann}}[\rho] = \mbox{Tr}(\rho \mbox{log}\rho)$ 
for $\rho$ the QM density matrix.
These notions have suitable properties and remain applicable \cite{arxiv} in passing to QFT and GR contexts. 
One contention in interpreting (0) at the general level required for developing a POT strategy is that 
information is minus entropy and classical (never mind quantum) gravitational entropy is a concept that 
is not well understood or quantified for general spacetimes \cite{arxiv}.  
Quantum gravity may well have an information notion 
$
I[\rho_{\mbox{\scriptsize QGrav}}] = 
\mbox{Tr}\rho_{\mbox{\scriptsize QGrav}}\mbox{log}\rho_{\mbox{\scriptsize QGrav}}, 
$
but either the quantum-gravitational density matrix is an unknown object since the  
underlying microstates are unknown, or, alternatively, one would need to provide an extra procedure for 
obtaining this, such as how to solve and interpret the WDE, which would be fraught with numerous further 
technical and conceptual problems.

Rather than a notion of gravitational information that is completely general, a notion of entropy 
suitable for approximate classical and quantum cosmologies may suffice for the present study. 
Quite a lot of candidate objects of this kind have been proposed.  
However, it is unclear how some of these would arise from the above fundamental picture, while 
for others it is not clear that the candidate does in fact possess properties that make it a bona fide 
entropy \cite{arxiv}.  
%
%Monotonicity is one often-mentioned property (with which gravitational information candidates based on 
%the Weyl tensor \cite{Weyls} have run into problems), 
%
%while information/entropy is characterized by a number of further properties \cite{Wehrl} that it is not 
%clear that the gravitational candidates have been screened for.  
%
Cosmologically relevant information notions proposed to date include some that are manifestly related 
to the above conventional notions of information, and also \cite{HBM} use 
$
I_{\mbox{\scriptsize HBM}}[\varepsilon] = 
\int\d\Omega \varepsilon \mbox{log}({\varepsilon}/{\langle\varepsilon\rangle}) 
$
and 
$
I^{\prime}_{\mbox{\scriptsize HBM}}[\varepsilon] =  
\langle\varepsilon \mbox{log}({\varepsilon}/{\langle\varepsilon\rangle})\rangle$, 
the first of which is a relative information type quantity (see Sec 6).

B) However, whether there is a pattern in a record or collection of records (and whether 
that pattern is significant rather than random) involves more than just how much information is 
contained within.  
Two placings of the same pieces on a chessboard could be, respectively, from a grandmasters' game and 
frivolous.  
What one requires is a general quantification of there being a pattern.  
This should be linked at least in part to information content, in that the realization of at least some 
complicated patterns requires a minimum amount of information.     
Records Theory is, intuitively, about drawing conclusions from similar patterns in different records.

Consider also the situation in which information in a curve or in a wave pulse that is  
detectable by/storeable in a detector in terms of approximands or modes.  
As regards localized useable information content per unit volume, considering the Joos--Zeh dust--CMB  
and $\alpha$-track--bubble chamber side by side suggests that most records in nature/one's model will be 
poor or diffuse.
For the Joos--Zeh \cite{JZ} example the `somewhere' is all over the place: 
``{\it in the vastness of cosmological space}".   
Detectors, such as the extension of Halliwell's 1-piece detector model (Sec 3.3, \cite{H99})
to a cluster, could happen to be tuned to pick up the harmonics that are principal contributors in the signal.  
In this way one can obtain a good approximation to a curve from relatively little information.  
E.g. compare the square wave with the almost-square wave that is comprised of the first 10 harmonics 
of the square wave.  
That is clearly specific information as opposed to information storage capacity in general.  
Likewise, a bubble chamber is attuned to seeing tracks, a detector will often only detect certain 
(expected) frequencies.  
Through such specialization, a record that `stands out' can be formed.    
One should thus investigate is quantitatively which of the $\alpha$-track and `dust grain' paradigms 
is more common.

C) Information can be lost from a record `after its formative event' -- the word ``stored" in (0) 
can also be problematic.     
Photos yellow with age and can be defaced or doctored.

%========================================================================================================
%========================================================================================================
%========================================================================================================
\section{Correlations between records}
%========================================================================================================
%========================================================================================================
%========================================================================================================

One concept of possible use is {\it mutual information}: this is a notion $M(A, B)$ 
$= I(A) + I(B) - I(AB)$ for AB the joint distribution of A and B for each of classical Shannon or 
QM von Neumann information.  
This is a quantity of the {\it relative information type} \cite{arxiv},    
$I_{\mbox{\scriptsize relative}}[p, q]         = \sum_x p_x\mbox{log}({p_x}/{q_x})$ 
(discrete case), $I_{\mbox{\scriptsize relative}}[\sigma, \tau] = \int \d\Omega \sigma\mbox{log}
({\sigma}/{\tau})$ (continuous case), 
(the object in Sec 4.1 is a special case of the continuous case of this in which the 
role of the second distribution is played by the average of the first).  
The QM counterpart of relative information is 
$I_{\mbox{\scriptsize relative}}[\rho_1, \rho_2] = \mbox{Tr}(\rho_1\{ \mbox{log}\rho_1 - \mbox{log}\rho_2\})
$
; mutual information also has QM analogues.  
It is not clear that these notions cover all patterns.  
Two records could be part of a discernible common pattern even if their constituent information is 
entirely different, e.g. the pattern to spot on two chessboards could be interprotection, 
manifest between rooks on one and between knights on the other.

Another is the family of notions of correlator/n-point function in the cosmological or QFT 
senses (or both at once).

%========================================================================================================
%========================================================================================================
\section{Further features of Records Theory}
%========================================================================================================
%========================================================================================================

Barbour furthermore asks \cite{B94II} whether there are any {\it selection principles} for such records 
(which he calls `time capsules'; the bubble chamber with the $\alpha$-particle track within is a such).    
If these features are to be incorporated, one would additionally need a (relative) measure of semblance 
of dynamics.   
How does a record achieve this encodement? 
Are SC's that encode this generic?  
Let us suppose that this is actually a special rather than generic feature for a SC to 
have.  
This would be the case if the dust grain--CMB photon paradigm is more typical than the 
$\alpha$-particle--bubble chamber one.  
Then one would have the problem of explaining why the universe around us nevertheless contains a 
noticeable portion of noticeably history-encoding records, i.e. a selection principle would be needed.

Barbour suggests a selection principle based on the following layers \cite{B94II}. 
1) There are some distinctive places in the configuration space.
2) The wavefunction of the universe peaks around these places, making them probable.   
3) These parts of the configuration space contain records that bear a semblance of dynamics 
(`time capsules').
[Following my arguments in Sec 1, this should be rephrased in terms of SC's.]  
Some doubts are cast on this scheme in \cite{arxiv}.  
In particular, A) Barbour supplies no concrete mathematical model evidence for there being any correlation 
between SC's being time capsules and their being near a distinctive feature of 
SC space such as a change of stratum or a point of great uniformity.  
B) Semiclassicality might either explain or supplant Barbour's selection principle,  while 
there are additionally two further a priori unrelated selection principles in the literature, which 
could be viewed either as competitors or as features that Barbour's scheme should be checked to 
be able to account for: I) {\sl branching processes} and II) {\sl consistency conditions in the Histories 
Theory framework}.

One reason that Barbour favours the above scheme is so as to be open to the possibility of explaining 
the Arrow of Time, unlike I), II), \cite{CastAsym} (which builds in a time asymmetry in the choice 
of admitted solutions), and Page's scheme (which is subject to the difficulties pointed out in Sec 4.4).
These various interesting issues should be further investigable using RPM's.

As regards Records Theory as a POT resolution, limitations exposed in this seminar are as follows.    
Records are ``somewhere in the universe that information is stored when histories decohere".     
But a suitable notion of localization in space and in configuration space may be hard to come by and/or 
to use for quantum gravity in general -- `where' particular records are can be problematic to quantify, 
and the records can be problematic to access and use too, since the relevant information may be 
`all over the place'.  
Also, `information' is problematic both as it may be of too poor a quality to reconstruct the history 
and because a suitably general notion of information is missing from our current understanding of 
classical gravity, never mind quantum gravity with its unknown microstates (mechanical toy models are 
useful in not having this last obstruction).  
Finally, the further Records Theory notions of significant correlation patterns and how one is to deduce 
dynamics/history from them looks to be a difficult and unexplored area even in simpler contexts than 
gravitation.  

\mbox{ }

%====================================================================================================
%====================================================================================================
\noindent{\bf{Acknowledgments}}
%====================================================================================================
%====================================================================================================
%
\noindent I thank Peterhouse for funding and  the organizers of TAM 2007 for inviting me.

%=====================================================BIBLIOGRAPHY==========================================================================

\end{document}